# Multi-Wavelength Selective Thermal Emission Enabled by Dual-Layer Localized Surface Plasmon Polaritons

SHUANG PAN[1,2], SHAOTENG WU[2,3], HUIXUE REN[2], JIARONG ZHAO[2], YUANHAO ZHU[2], SAILEI LI[2,] LI HE[2], JUN-WEI LUO[2,4]

[1] *School of Microelectronics, University of Science and Technology of China, Hefei, Anhui 230026, People's Republic of China*

[2] *State Key Laboratory of Superlattices and Microstructures, Institute of Semiconductors, Chinese Academy of Sciences, Beijing 100083, People's Republic of China*

[3] *Email: wst@semi.ac.cn*

[4] *Email: jwluo@semi.ac.cn*

**Abstract:** Thermal emission is a ubiquitous electromagnetic wave with an extreme broad spectrum in nature, and controlling thermal emission can be used to develop low-cost and convenient infrared light sources with wavelength tunable in a wide range that is currently difficult to other sources. Conventional metasurfaces are commonly used to control light but lack the flexibility to achieve complex emission spectral profiles and dynamic tuning. Here, we introduce a novel dual-layer metasurface structure with two completely independent layers to achieve a multi-peak thermal emission within the 5-8 μm wavelength range. Simulations and experiments show that this two-layer structure can achieve arbitrary spectral shapes without interfering with multiple resonant modes. This unique configuration presents a promising platform for further exploration in thermal emission engineering, enabling spectral control and dynamic tuning.

## 1. Introduction

The development of advanced thermal emitters with tailored spectral and directional characteristics has garnered extensive attention due to their wide range of potential applications in fields such as energy harvesting [1], thermal imaging [2], and sensing[3]. However, conventional thermal emitters often suffer from low emissivity and limited control over the emitted spectrum, hindering their performance in these applications. Metasurfaces, composed of subwavelength structures engineered to manipulate light-matter interactions, have emerged as a promising platform for achieving unprecedented control over thermal emission. Metamaterial-based light manipulation has been extensively studied in recent decades, offering promising potential applications as it can replace bulky conventional optical components by modulating various properties of light [4, 5] through subwavelength structures. In metamaterial research, metals are frequently used materials. According to the Drude model, metals can achieve negative permittivity in the visible and infrared wavelengths due to their extremely high electron density. This enables the emergence of surface plasmon polaritons (SPPs) at the interface with ordinary dielectrics. Under specific conditions, these SPPs can resonate with free-space electromagnetic waves, thereby achieving prominent emissivity/absorptivity of certain wavelengths. This ability to control light has also opened new avenues in the field of thermal emission [6-8]. For instance, potential applications includes naked-eye 3D displays [9], optical cloaks [10], and research in fields such as light-emitting diodes (LEDs) [11, 12] and lasers [13, 14]. By carefully designing the geometry and arrangement of these subwavelength structures, metasurfaces have been demonstrated to be able to enhance the emissivity, generate linearly polarized light [15], direct the emitted radiation [7, 16-18], and even shape the spectral profile of the emitted light [15, 19, 20].

However, conventional metasurfaces often lack the flexibility to achieve complex spectral profiles and dynamic tuning. For instance, the single-layer resonance implemented by Ye *et al.* can realize two emission peaks at 6.3 μm and 10.5 μm through different modes of the same grating [15], but the relative positions between the two peaks are difficult to be adjusted freely. Huang *et al.* has utilized the refractive index of different materials to achieve multi-peak emission of mixed medium materials [21] to introduce additional flexibility, but itis still a single-layer structure and, thus does not allow for completely independent control of each peak. Chen *et al.* has proposed dual-layer structures to further tune the wavelength of the thermal emission [22]. However, it can obtain limited regulation capabilities since it is only be adjusted by controlling the type and thickness of the medium materials as a result of the width and period of the upper and lower layers must remain consistent.

To overcome these limitations to have a higher degree of adjustability, in this work, we propose a novel dual-layer metal-insulator-metal (MIM) metasurface thermal emitter with the two metal layers being spatially independent via separately setting their period and grating width. We demonstrate that this structure can achieve an independent shift of each emission peak within the 5-8 μm band. This structure can be applied in gas sensing, by designing the structure, it is possible to emit and absorb light of different wavelengths from the same light source point, enabling simultaneous monitoring of different gases. Additionally, in the field of 3D displays, this structure also holds great potential, as it can independently emit different polarized light and wavelengths from the same pixel. This can significantly reduce the size of 3D display devices, the structure also has a high degree of compatibility with voltage control, as different voltages can be applied independently to different layers, offering broad prospects for future applications. This design brings thermal emission regulation to a new level of flexibility.

## 2. Structure construction and optical testing

The dual-layer MIM structure is designed to emit multi-wavelength thermal emission when placed on a custom-built hot plate, as the schematic picture illustrated in Figure 1. The structure consists of a 100 nm thick $SiO_2$ layer deposited on top of the silicon substrate to prevent leakage current during future voltage application, followed by a deposition of a 200 nm thick Al layer, which serves as the reflector, and finally a 300 nm thick $SiO_2$ layer acting as the thermal emission source. Note that the 300 nm $SiO_2$ layer can also be replaced with other dielectrics to tune the wavelength of the emission peak by changing the dielectric constants. The unique feature of this thermal emission source is that it consists of two distinct grating layers. The first grating (FG) layer consists of 5/90/5 nm Ti/Ag/Ti and is embedded within the 300 nm $SiO_2$ layer, and the second grating (SG) is on top of it with 10/100 nm Ti/Ag. In each grating layer, the Ti layer acts as an adhesive layer to enhance the adhesion between Ag and $SiO_2$ [15]. In addition to metals, highly doped semiconductors can also be used as reflectors to provide better tunability with the sacrifice of less reflection [23].

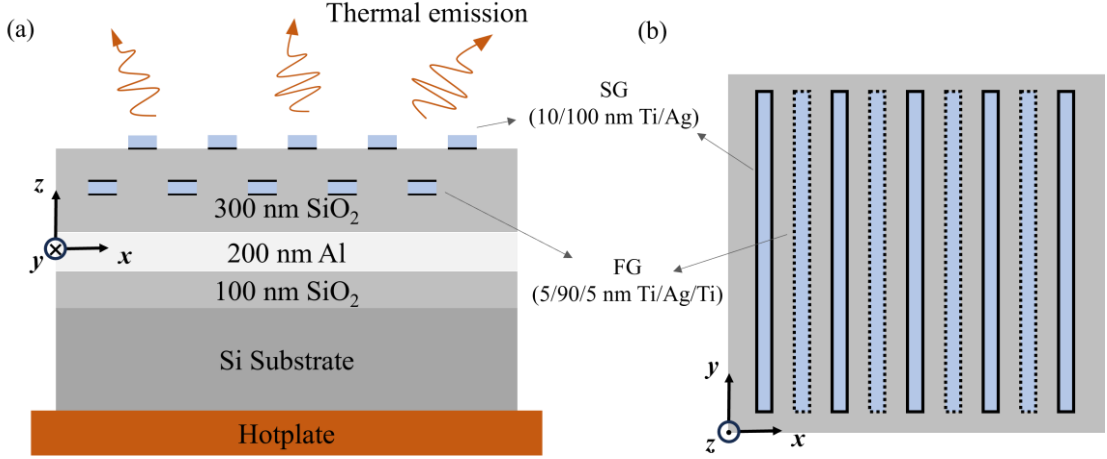

Figure 1. (a) Cross-section and (b) top view of the dual-layer metasurface structure. The structure consists of a 100 nm thick SiO$_2$ layer deposited on top of the silicon substrate to prevent leakage current during future voltage application, followed by a deposition of a 200 nm thick Al layer, which serves as the reflector, and finally a 300 nm thick SiO$_2$ layer acting as the thermal emission source. The first grating (FG) layer consists of 5/90/5 nm Ti/Ag/Ti and is embedded within the 300 nm SiO$_2$ layer, and the second grating (SG) is on top of it with 10/100 nm Ti/Ag.

The thermal light emission with specific wavelengths from this novel structure is due to the resonance with the SPPs [24, 25] and the localized surface plasmon polaritons (LSPPs) [15, 20]. To achieve a resonance of SPPs with light requires momentum-matching techniques, such as periodic disks [26-28], gratings [15], crosses [29], holes [25], etc. The one-dimensional grating needs to satisfy the grating phase matching condition $\bm{k_{sp} = k_x + iG}$, where $|k_x| = |k_0|sin\theta$ ($k_0$ is the wavevector of the incident light with an incident angle $\theta$), $|G| = 2\pi/a$ is the reciprocal lattice vector with $a$ being the grating period, and $i$ is an integer [25]. It leads to a strong dependence of the emissivity on the period $a$ and angle $\theta$, which can be adjusted to achieve a high spatial coherence. Furthermore, since surface plasmon polaritons can only exist in the transverse magnetic (TM) mode [30], unpolarized light can be converted into polarized light as long as the structure is designed to to break the 90-degree rotational symmetry in the x-y plane. On the other hand, the resonance of the LSPPs with the light occurs in the cavity formed by the impedance mismatch between the region beneath the grating and the adjacent region [31]. The resonant wavelength $\lambda$ has to satisfy the standing wave condition: $2d = m\frac{\lambda}{n_{eff}}$, where $d$ is the grating width, $m$ is a positive integer, and $n_{eff}$ is the effective refractive index in the resonant cavity [24]. Moreover, its group velocity $\frac{\partial \omega}{\partial k_x} = 0$ indicates that the LSPPs resonant mode is angle-insensitive, resulting in a wide absorption/emission angle [15, 20, 22, 28, 32]. This wide-angle emission/absorption feature of the LSPPs makes it easy to distinguish them from the SPPs.

The Finite-Difference Time-Domain (FDTD) method is utilized to simulate the light emission of this structure. According to the Kirchhoff's law, the absorptivity equals the emissivity (α=E) [6]. Due to the presence of a thick Al layer (~200 nm) as a reflector, the transmissivity can be considered negligible (T≈0). Therefore, the reflectivity is given by $R = 1 - \alpha = 1 - E$, implying that the thermal emissivity could be obtained from the corresponding reflectivity [19]. We perform the simulation for a structure containing two grating layers with periods of 10 µm: FG has a fixed width of 2.5 µm and the width of SG varies from 1.5 to 5 µm. Figure 2 (a) shows that the simulated reflectivity spectrum ranges from 5 to 9 µm for the incident light being set to be linearly polarized with the electric field along the x-axis. Since the simulated SG width is discrete, there are some saddle points, which are labeled in Figure 2 (a), which do not affect the analysis of the results. Figure 2 (a) verifies that this structure can independently control the movement of each peak and has high flexibility. It can be seen that the minimum value of the reflectivity at 7.24 µm does not change with the SG width due to it is the resonance mode related to the FG. Figure 2 (a) also shows that the increase of the SG width causes the redshift of two modes, which are the first-order (m=1) and third-order (m=3) LSPPs modes related to the SG, respectively. The orders of the modes can be verified by observing the distribution of the magnetic field intensity of corresponding resonant wavelength, as shown in Figure 2 (b), (c) and (d). Figure 2 (b), (c) and (d) respectively show the magnetic field distributions of the 2.5µm FG and 5µm SG structures at wavelengths of 7.24µm, 7.80µm, and 5.52µm.

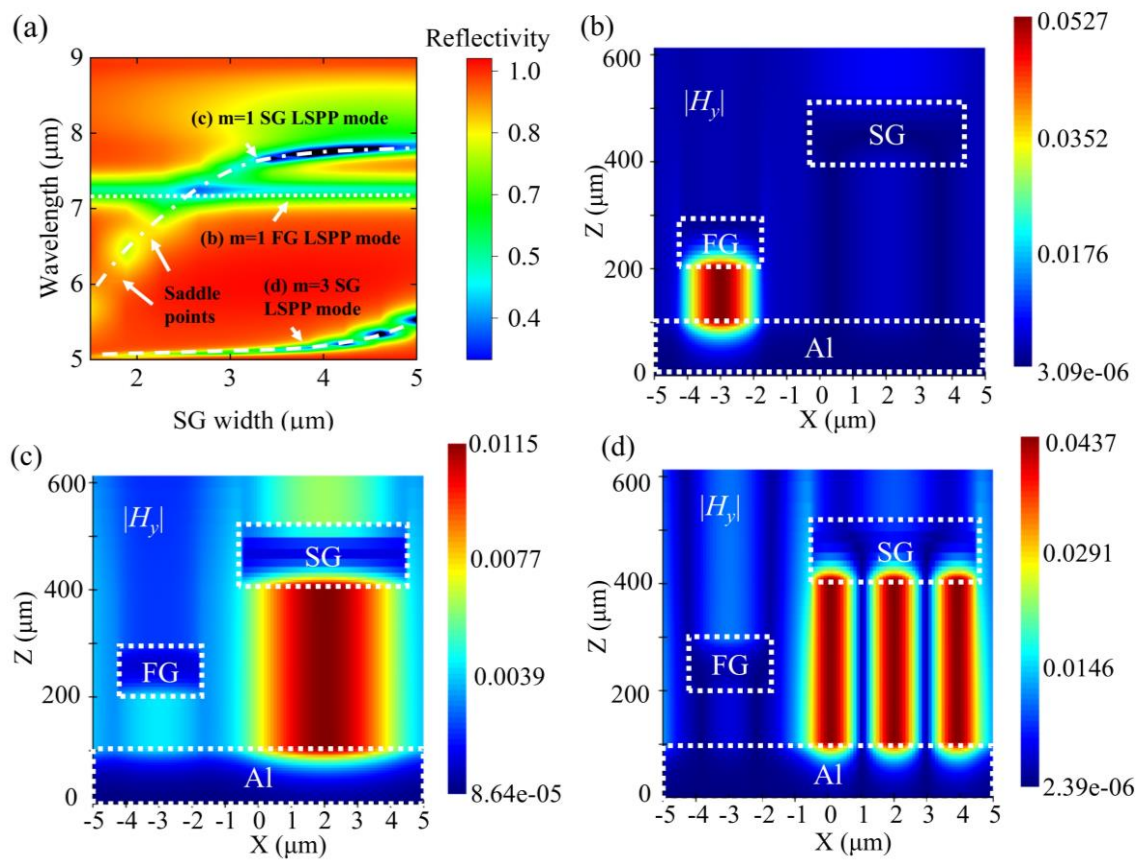

Figure 2. (a) A map of reflectivity as a function of SG width and wavelength. The magnetic field distribution at wavelengths of (b) 7.24 µm, (c) 7.80 µm and (d) 5.52 µm for the structure with 2 µm FG and 5 µm wide SG. (a) shows that the magnetic field is confined below the FG, corresponding to the m=1 LSPPs mode of the FG. (c) and (d) correspond to the m=1 and m=3 LSPPs resonant modes of the SG, respectively.

According to the design structure and simulation results, concrete samples were made. The fabrication process of this designed dual-layer grating structure presents a challenge in embedding the FG within the $SiO_2$ layer while maintaining a relatively flat surface on top. This was achieved through a multi-step process, as illustrated in Figure 3. The process began with a clean of a silicon wafer serving as the substrate followed by a deposition of a 100 nm thick $SiO_2$ insulating layer onto it using plasma-enhanced chemical vapor deposition (PECVD). This was then followed by the deposition of a 200 nm thick Al reflective layer via electron beam evaporation and an additional 200 nm thick $SiO_2$ layer using PECVD. To enhance the density of the deposited $SiO_2$ layer, the sample in the furnace was gradually heated from room temperature to 400 °C over one hour, then held at 400 °C for two hours, and finally allowed natural cooling to room temperature. Next, a negative photoresist was utilized to define a series of grating patterns with a period of 10 μm and a width of $w_1$. The photoresist was then baked in an oven (room temperature to 120 °C over 40 minutes, held at 120 °C for 40 minutes) instead of a hot plate to ensure etch resistance and successful lift-off. Reactive ion etching (RIE) was employed to etch 100 nm of $SiO_2$, creating grooves with a width of $w_1$. The FG was then formed through the deposition of a 5 nm/90 nm/5 nm Ti/Ag/Ti stack using electron beam evaporation followed by lift-off. Subsequent deposition of $SiO_2$ using PECVD was followed by the application of a negative photoresist to define the SG pattern with a period of 10 μm and a width of $w_2$. Finally, 10 nm Ti and 100 nm Ag were deposited, and a lift-off process was employed to form the SG [22]. The resulting pattern of FG and SG occupied a square of 0.36 cm$^2$.

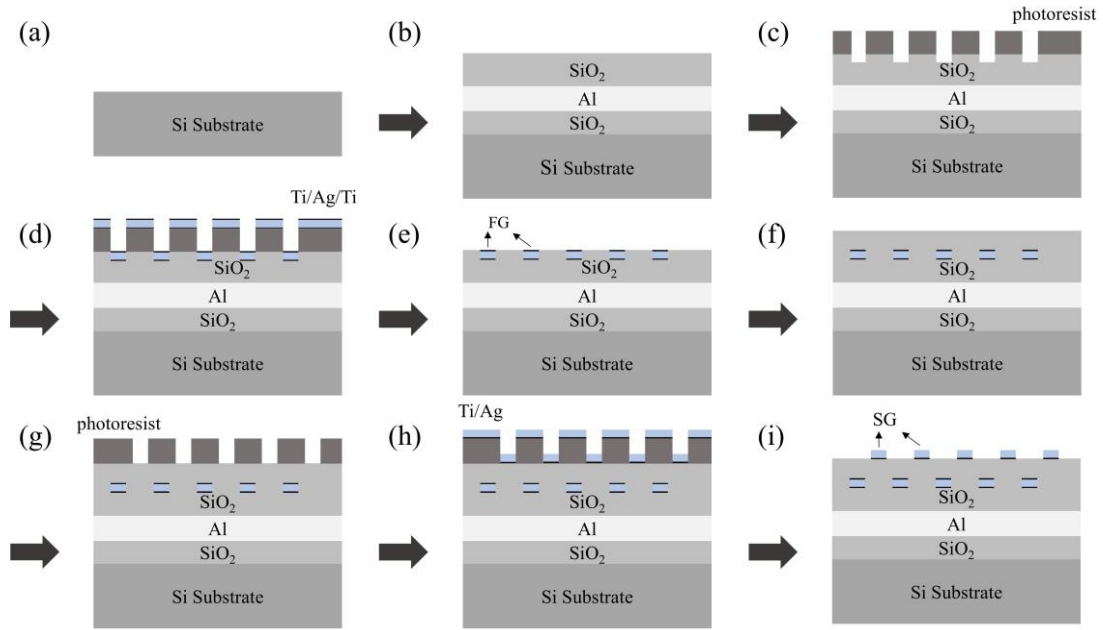

Figure 3. A series of sectional diagrams are used to show the steps of manufacturing the sample.

Four samples, numbered #1 to #4, were fabricated using the procedure mentioned above. The FG width is fixed at 2.47 μm, and the SG widths are 2.52, 3.76, 4.52, and 5.67 μm, respectively, for samples #1 to #4. The actual geometric parameters of the samples are listed in Table 1. Figure 4 (a) shows a cross-sectional SEM image of the fabricated sample #1, and Figure 4 (b) is a magnified view of Figure 4 (a). The image clearly reveals the presence of two grating layers, with their geometrical dimensions closely matching the design specifications. Notably, the $SiO_2$ top surface exhibits minimal unevenness. This observation aligns well with the design expectations and confirms the successful fabrication of the sample.

**Table 1. Geometrical parameters of the four samples**

| Sample | $w_1/\mu m$ | $w_2/\mu m$ |
|---|---|---|
| 1 | 2.47 | 2.52 |
| 2 | 2.47 | 3.76 |
| 3 | 2.47 | 4.52 |
| 4 | 2.47 | 5.67 |

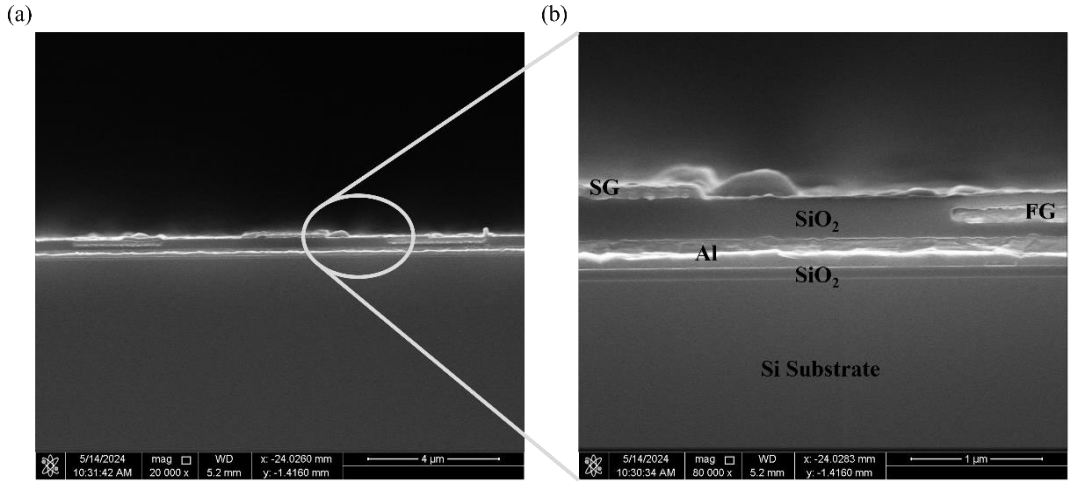

Figure 4. (a) SEM cross-section of Sample #1. (b) is a magnified view of (a).

Thermal emission spectra of the fabricated samples were measured using a Fourier Transform Infrared (FTIR) system equipped with a Mercury Cadmium Telluride (MCT) detector and a KBr beamsplitter. Different temperatures of thermal emission were achieved by placing the sample on a custom-built hot plate. A silicon wafer coated with blackbody paint and an unpatterned sample of 200/300 nm $Al/SiO_2$ were utilized for analysis. Both samples were 1.44 $cm^2$ squares. The net emissivity due to the grating was obtained by subtracting the spectrum of the unpatterned reference sample from the thermal emission spectrum of our sample and then dividing the result by the spectrum of the black paint sample: $E = \frac{E_{device} - E_{blank}}{E_{black}}$. Figure 5 (a) displays the emissivity measured at 310°C compared to the spectrum simulated using the Finite-Difference Time-Domain (FDTD) method based on the geometrical parameters observed from SEM, as depicted in Figure 5 (b). Note that its ordinate is 1-R to make it more convenient to compare with the emissivity.

In Figure 5 (a) and (b), the experimental and simulation results are observed to closely align, and each peak is marked with corresponding resonance mode. Due to the fixed widths of the FG of four samples, resonant peaks are observed at 7.2 μm for all samples (At the dotted line in Figure 5 (a) and (b)), these peaks are due to the first-order LSPPs resonance of FG. In contrast, the first and third-order LSPPs modes of SG redshift with increasing SG width. This demonstrates a high level of flexibility, allowing independent control of the wavelength of each peak through structural adjustments.

Furthermore, the full width at half maximum (FWHM) in the experimental results is larger, which is due to the additional scattering caused by the fact that the material interface in the actual fabrication process is not perfectly flat. Further optimization of the fabrication process could potentially improve this aspect. Among the four samples, only Sample 4 shows a good match between experimental and simulated results for the third-order LSPPs resonance peak of the SG. This discrepancy of other samples is speculated to be influenced by the absorption peak of $CO_2$ at 4.3 μm. Since the emissivity spectrum is obtained by subtracting the emission intensity of the reference sample from that of the samples and then dividing the results by the emission intensity of a blackbody, the minimum value of the emission intensity of a blackbody caused by the absorption of $CO_2$ leads to significant fluctuations after division. Therefore, the emissivity results near 4.3 μm are not reliable. In short, the experimental results basically agree well with the simulation results, which verify that the simulated peak can move independently, and realize a more flexible means of light regulation.

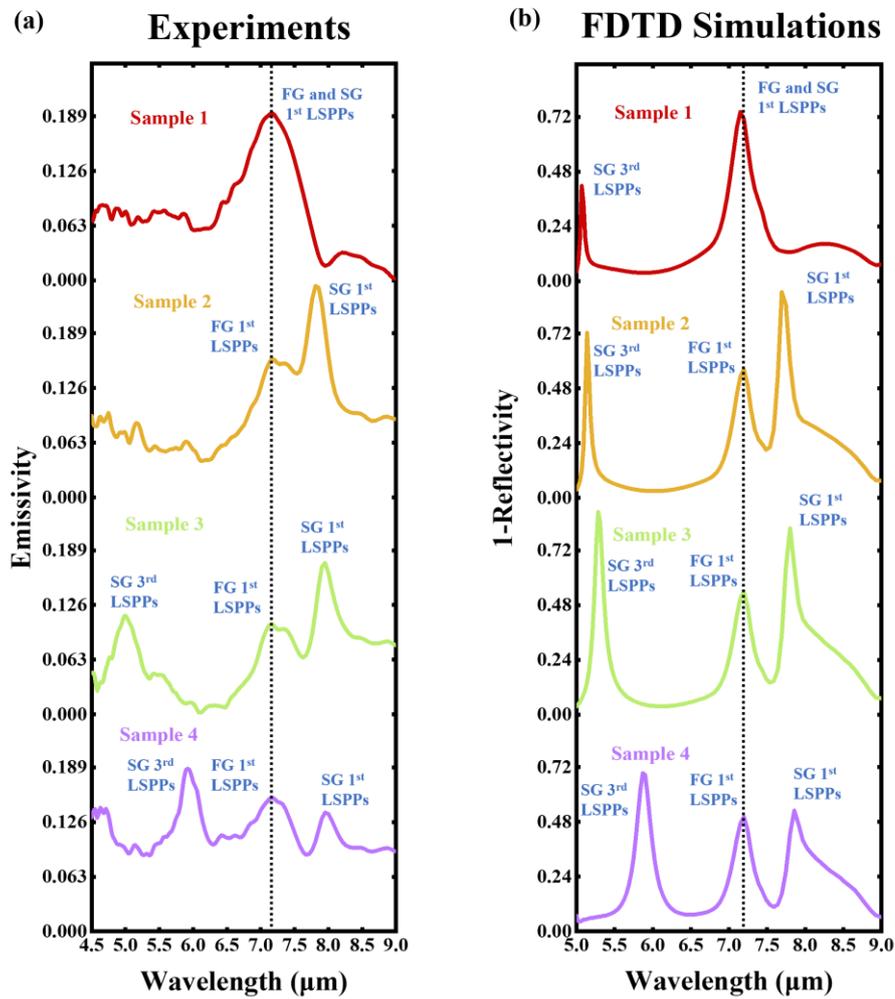

Figure 5. (a) The emissivity spectra of four samples measured experimentally ; (b) The corresponding simulation predicted reflectivity spectra based on the geometrical parameters obtained from the SEM cross-section of four fabriacated samples (the ordinate is 1-R).

**3. Detail simulation and discussion**

This section simulated the optical emission characteristics of the structure in more detail using the FDTD method and explained some phenomena of the results. As a representative example, the electromagnetic field distribution at a wavelength of 5.52 μm was simulated using a structure with 2 μm FG and 5 μm SG. The resonant cavity of this structure only allows for electromagnetic fields with $H_y$ and $E_z$ components, as shown in Figure 6 (a) and (c). Figure 6 (b) shows that the $E_z$ component is almost zero under the grating. Due to the polarization direction of the incident light, the $H_x$, $H_z$, and $E_y$ components are zero in the entire space (not shown). Therefore, if the incident light only has $H_x$ and $E_y$ components, the obtained reflectivity in the infrared band will be almost equal to 1. These conclusions can be easily derived from Maxwell's equations. Due to the large imaginary part of the metal's dielectric function, the electromagnetic field penetration depth is very small, allowing us to approximate the metal interior as field-free. Due to the resonant cavity thickness is much smaller than the wavelength, the electromagnetic field inside the cavity can be considered uniform along the z direction. This facilitates the derivation of $E_x$, $E_y$, and $H_z$ being 0 from the boundary conditions $\boldsymbol{n} \times (\boldsymbol{E_1} - \boldsymbol{E_2}) = 0$ and $\boldsymbol{n} \cdot (\boldsymbol{B_1} - \boldsymbol{B_2}) = 0$. Applying $\nabla \times \boldsymbol{E} = -\frac{\partial \boldsymbol{B}}{\partial t}$ and treating $E_z$ as solely a function of $x$, then $H_x$ is also 0. Consequently, this resonance mode results in emitting polarized light or absorbing only certain polarization component of incident light [15].

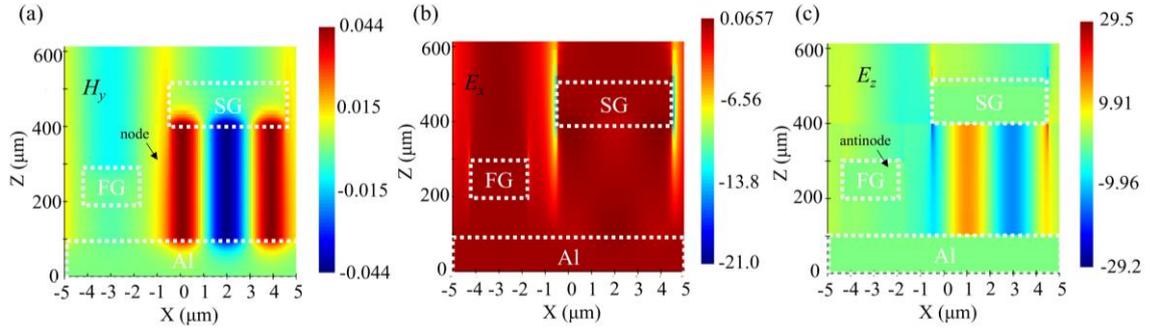

Figure 6. The distribution of (a) Hy, (b) Ex, and (c) Ez components for a structure with 2 μm FG and 5 μm SG at a wavelength of 5.52 μm.

Three novel phenomena are observed in the simulation results of this structure. The first one is that the magnetic field node and the electric field antinode of the standing wave are always located at the edge of the resonant cavity, as shown in Figure 6 (a) and Figure 6 (c), respectively. This phenomenon arises from a combination of the resonant cavity is mirror-symmetrical, and the standing wave is composed of electromagnetic waves entering from both ends of the grating. Since the structure does not cause scattering of the magnetic field, the magnetic field direction of two incoming waves remains along the *y* direction with equal magnitude. However, the propagation directions (which is also the direction of $k$) of the two incoming waves are opposite. According to $\boldsymbol{k} = \omega \boldsymbol{E} \times \boldsymbol{B}$, the direction of $\boldsymbol{E}$ must be opposite. Therefore, when these two light beams are superimposed, the resulting electric field is always zero in the middle of the grating, and the nodes of the standing electric wave are at both ends of the grating in the odd-order modes (m is odd). The second one is that only the odd-order modes can be observed in our simulation results, as shown in Figure 2 (b), (c) and (d). Under the condition that the incident angle of the incident light is set to zero during simulations, symmetry enforces the net electric dipole vanishing on the grating for even-order modes, so it won't couple with the electric field. However, for odd-order modes, a net electric dipole can always exist. Hence we explain the absence of even-order

LSPPs modes under normal incidence, as demonstrated by Y. Todorov *et al* [31]. Finally, the SG resonance peak remains nearly unchanged when the wavelength approaching 8 μm, as shown in Figure 2 (a). It can be explained by the refractive index, based on the resonance formula $2d = m\frac{\lambda}{n_{eff}}$, the effective refractive index for various wavelengths in the resonant cavity can be easily calculated, as shown in Figure 7. It can be seen that when the wavelength approaches 8 μm, the effective refractive index drops sharply, making it difficult for SG modes to continue to redshift at wavelengths close to 8 μm. Comparing effective refractive index with the refractive index of the $SiO_2$ thin film, it is observed that the effective refractive index is only 1.4 times that of $SiO_2$, so the decrease here comes from the asymmetric longitudinal optical phonon resonance absorption of $SiO_2$.

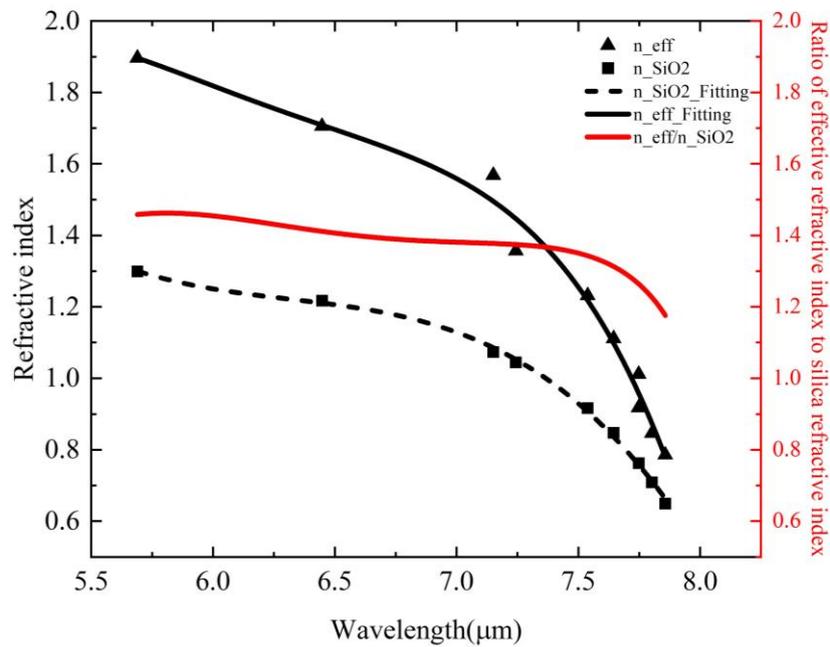

Figure 7. Comparison of effective refractive index and $SiO_2$ refractive index of thin film under metal grating.

Nevertheless, the previous results did not exhibit any evidence of a coupled SPPs mode in the simulated spectrum. It has been established earlier [25] that the formation of SPPs is highly dependent on the grating period, while LSPPs are less sensitive to period variations. To enhance the visibility of the SPPs mode, we conducted additional simulations with an identical grating width but varying periods. For clarity, the FG structure was excluded, leaving only the SG with a width of 5 μm. Figure 7 (a) reveals an SPPs resonance mode within the wavelength range of 5 to 7 μm. At a period of 10 μm, the SPPs and LSPPs modes nearly coincide. When this coincidence occurs, the energy is predominantly absorbed by LSPPs, rendering the SPPs mode indiscernible in Figure 2. Figure 8 (b) depicts the magnetic field distribution at the absorption peak of 6.42 μm wavelength and a period of 12.67 μm. This distribution of Figure 8 (b) exhibits a distinct departure from the standing wave mode observed in Figure 2 (b), (c) and (d), confirming the presence of SPPs.

We have elucidated the formation and distribution of SPPs and LSPPs resonance modes using a dual-layer grating structure. This structure offers significant flexibility for precisely manipulate the LSPPs and SPPs through tunning the width and period of the grating layers. This dual-layer structure not only provides enhanced versatility in static structural design but also presents a promising platform for dynamic voltage regulation of multiple emission peaks [4, 33-36].

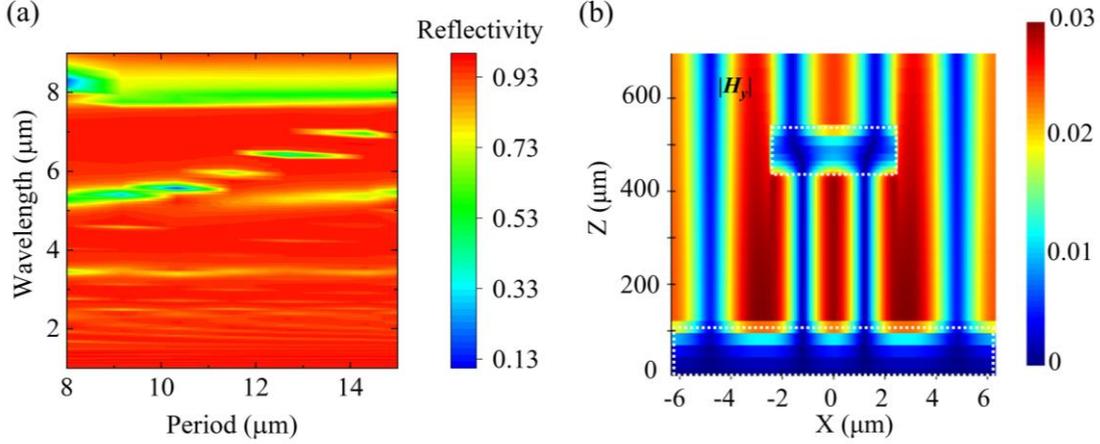

Figure 8. (a) A map of reflectivity as a function of period and wavelength. (b) Magnetic field intensity distribution at 6.42 μm wavelength and 12.67 μm period.

## 4. Conclusion

This work presents a comprehensive investigation of a dual-layer grating structure for controlling thermal emission, encompassing both simulation and experimental verification. The key findings demonstrate the independent nature of the two grating layers, confirming that the LSPPs resonance modes of the FG and SG do not exhibit mutual interference. We further demonstrate a direct positive correlation between grating width and the wavelength of the LSPPs resonance mode. Notably, the resonant cavity formed beneath the grating supports only odd-order standing wave modes, such as the first and third orders. Our analysis reveals that metallic gratings increase the refractive index of $SiO_2$ by 1.4 times, while the refractive index at the wavelength of absorption peak of $SiO_2$ LO phonon decreases significantly, causing it difficult for LSPPs modes to continue to redshift at wavelengths close to 8 μm. Furthermore, the simulation elucidates the phenomenon of third-order mode overlap with SPPs energy when the period is 10 μm, resulting in the suppression of SPPs resonance. However, if the resonance energy of SPPs is deviated from the resonance energy of LSPPs by adjusting the period, the resonance of SPPs can be observed.

The proposed structure significantly enhances the flexibility of light control. It not only enables thermal emission at specific wavelengths but also allows for independent movement of individual thermal emission peaks, even enabling the construction of arbitrary wavelengths. It holds promise for applications in low-cost infrared light sources and provides valuable insights for designing advanced light source control devices. This structure also has potential applications in fields such as gas sensing,

3D displays, and various other areas. Future research will focus on exploring the dynamic tuning capabilities of this system, unlocking its full potential for advanced thermal emission applications.

**Funding.** This work was support by the National Natural Science Foundation of China (62474176), Beijing Natural Science Foundation under Grant 4242061, the Key Research Program of Frontier Sciences, CAS, under Grant No. ZDBS-LY-JSC019, CAS Project for Young Scientists in Basic Research under Grant no. YSBR-026, the Strategic Priority Research Program of the Chinese Academy of Sciences under Grant No. XDB43020000, and Joint Fund of Henan Province Science and Technology Research and Development Program under Grant 225200810014.

**Disclosures.** The authors declare no conflicts of interest.

**Data availability.** The data that support the findings of this study are available from the corresponding author upon reasonable request.